\documentclass[reviewcopy]{elsarticle}

\usepackage[reviewcopy]{adndt}
\usepackage{longtable}

\usepackage{mathptmx}

\usepackage{amsmath}

\usepackage{amssymb}

\biboptions{square,sort&compress}
\bibpunct[]{[}{]}{,}{n}{}{;}
\begin{document}

\begin{frontmatter}

\journal{Atomic Data and Nuclear Data Tables}


\title{Radiative rates for E1, E2, M1, and M2 transitions in S-like to F-like tungsten ions (W~LIX to W~LXVI) }

  \author[One]{Kanti M. Aggarwal\fnref{}\corref{cor1}}
  \ead{K.Aggarwal@qub.ac.uk}

  \author[One]{Francis P. Keenan}


  \cortext[cor1]{Corresponding author.}

  \address[One]{Astrophysics Research Centre, School of Mathematics and Physics, Queen's University Belfast,\\Belfast BT7 1NN,
Northern Ireland, UK}


\date{16.12.2002} 

\begin{abstract}  
Calculations of energy levels, radiative rates and lifetimes are reported for eight ions of tungsten, i.e. S-like (W~LIX) to F-like (W~LXVI). A large number of levels has been considered for each ion and  extensive configuration interaction has been included among a range of configurations. For the calculations,  the general-purpose relativistic atomic structure package ({\sc grasp}) has been adopted, and radiative rates (as well as oscillator strengths and line strengths) are listed  for all E1, E2, M1, and M2 transitions of the ions. Comparisons have been made with  earlier available experimental and theoretical energies, although these are limited to only a few levels for most ions. Therefore for additional accuracy assessments, particularly for energy levels, analogous calculations have been performed with the flexible atomic code ({\sc fac}). \\ \\
{\em Received}: 28 September 2015, {\em Accepted}: 12 February 2016

\vspace{0.5 cm}
\end{abstract}

\end{frontmatter}




\newpage

\tableofcontents
\listofDtables
\listofDfigures
\vskip5pc


\section{Introduction}

Tungsten (W) is  one of the most important constituents of tokamak reactor walls \cite{putt}. Additionally, it  radiates strongly over almost all ionisation stages. For example,  the most intense emission lines of W ions \cite{putt} are from W~XXII to W~L in  the  VUV  to  the  soft  x-ray  region,  covering  an
electron temperature range from about 0.5 to 5.0 keV. Similarly,   P{\"u}tterich et al.~\cite{putt} have predicted  emission features from  W~LXI to W~LXIX in the 
0.1--0.15 nm, 1.8--4.0 nm and around 8 nm ranges. However, to assess  radiation loss and for modelling plasmas, atomic data (including energy levels and  oscillator strengths or radiative decay rates) are required for many of the W ions. Their need for atomic data for several ions, including those of W, has increased significantly due to the developing  ITER project. Therefore, several groups of people are actively engaged in producing atomic data.

Early calculations for a number of W ions (W~XXXVIII to W~XLVIII) were performed by Fournier \cite{kbf}. He adopted a relativistic atomic structure code,  but reported only limited results for energy levels and oscillator strengths ($f$-values). A thorough critical compilation of experimental, theoretical and analytical energy levels of W ions (W~III through W~LXXIV) has been undertaken by Kramida and Shirai \cite{ks1} and has been further reviewed by Kramida \cite{ks2}. These energy levels, along with some spectral lines, are also available on the NIST (National Institute of Standards and Technology)  website at {\tt http://www.nist.gov/pml/data/asd.cfm}. Recently, spectra in the EUV wavelength range (4--20 nm) have been measured by Ralchenko et al. \cite{yuri}, for a number of W ions, namely W~LV to W~LXIV. Similarly, Clementson et al. \cite{ll1} have discussed spectroscopy of many W ions (W~XLVII to W~LXXII). On the other side, calculations have been performed for several W ions, such as by Quinet \cite{pq} for W~XLVIII to W~LXII. Although he adopted the {\sc grasp} code for the calculations, his reported results for energy levels and radiative rates ($A$-values) are confined to forbidden lines within the 3p$^k$ and 3d$^k$ configurations. However, for the modelling of plasmas, atomic data among a wider range of levels/transitions are preferred. Therefore, we have already reported such data for two W  ions, namely W~XL \cite{w40a,w40b} and W~LVIII \cite{w58a,w58b}. In this paper, we extend our work to eight other  W ions,  S-like (W~LIX) to F-like (W~LXVI).

As in our earlier research \cite{w40a,w40b,w58a,w58b} and those of others \cite{pq,ajm},  we have adopted the  fully relativistic multi-configuration Dirac-Fock (MCDF) atomic structure code  \cite{grasp0}, better known as the general-purpose relativistic atomic structure package ({\sc grasp}) \cite{grasp}. This code is  based on the $jj$ coupling scheme,  includes higher-order relativistic corrections arising from the Breit interaction and QED (quantum electrodynamics) effects, and is suitable for the heavy ions considered here.  However, this original version \cite{grasp0}  has undergone several revisions, such as by  \cite{grasp,grasp2k, grasp2kk}, and the one employed here (and by many other workers) has been revised by Dr. P. H. Norrington, and is freely available at  {\tt http://web.am.qub.ac.uk/DARC/}.

\section{Energy levels}

Extensive configuration interaction  (CI)  has been incorporated in GRASP, as described below for each ion, and  for the optimisation of the orbitals the option of  `extended average level' (EAL),  in which a weighted (proportional to 2$j$+1) trace of the Hamiltonian matrix is minimised, has been adopted.   The GRASP code has a few other choices for optimisation, such as average level (AL) and extended optimal level (EOL). However, in general,  the results obtained with the AL option are comparable with those of EAL as already discussed and demonstrated by us for several other ions, such as  those of Kr \cite{kr} and Xe \cite{xe}. Similarly, the EOL option may provide slightly more accurate data for a few predefined levels, but is only useful if the experimental energies are known, which is not the case for a  majority of the levels of the ions  studied here.

\subsection{S-like W~LIX}

Clementson and  Beiersdorfer \cite{cb} have measured wavelengths for  3 lines of W~LIX. They also calculated these with two different codes, i.e. {\sc grasp} and {\sc fac} (flexible atomic code), and there is no (major) discrepancy among the results.  For modelling purposes, Feldman et al. \cite{uf1} calculated atomic data for many W ions, including W~LIX, but did not report the data. Furthermore, they used a simple model consisting of the 3s$^2$3p$^4$,  3s3p$^5$,  3s$^2$3p$^3$3d, and 3p$^6$ configurations, generating 48 levels in total.

For our work, we have performed two sets of calculations using the {\sc grasp} code. In the first  (GRASP1) we have included 2762 levels of the all possible combinations of the $n$ = 3 orbitals, i.e.  18 configurations in number. The second  (GRASP2) involves an additional 28 configurations, which are [3s$^2$3p$^3$, 3s$^2$3p$^2$3d, 3s3p$^4$, 3s3p$^3$3d, 3s$^2$3p3d$^2$, 3s3p$^2$3d$^2$, and 3p$^5$]4$\ell$. These 46 configurations generate  12~652 levels in total. In Table~A we compare the energies obtained from both models, but for only the lowest 20 levels. Differences between the two sets of energies are less than 0.025 Ryd and the  inclusion of larger CI in the GRASP2 calculations has lowered the energies for most of the levels. Therefore, it is necessary to assess the effect of further CI on the energy levels. For this  we have adopted the {\sc fac} code of  Gu  \cite{fac}, which is also fully relativistic and is available from the website {\tt https://www-amdis.iaea.org/FAC/}. This code is comparatively more efficient to run and generally yields results similar to those obtained  with other atomic structure codes, as has already been demonstrated in several of our earlier papers -- see for example Aggarwal et al. \cite{fe15}. With FAC we have also performed two sets of calculations, i.e. FAC1:  includes the same 2762 levels as in GRASP1, and FAC2:  also includes levels of the 3$\ell^5$4$\ell$ configurations, generating 38~694 levels in total. Energies obtained from both these models are also listed in Table~A for  comparison.
 
Discrepancies between the GRASP1 and FAC1 energies are up to 0.15 Ryd (see level 13), in spite of including the {\em same} CI. This is because of the differences in the algorithms of the codes and  also in calculating the central potentials.  Additionally, the energies obtained from {\sc fac} are generally lower for most levels. However, inclusion of additional CI in the FAC2 calculations further lowers the energies, but only up to 0.02 Ryd for some of the levels. Therefore, it may be reasonable to say that the inclusion of CI in our GRASP2 calculations is sufficient to calculate accurate results, but differences with FAC2 remain of up to 0.15 Ryd.  The  NIST  compilation is  only for  a few levels of W~LIX, which are mostly based on the experimental and theoretical work of Clementson et al. \cite{ll1}. However, these energies are not very accurate as indicated on their website, and many levels are also missing from the compilation. Nevertheless, in Table~A we have included their energies for comparison. Unfortunately, differences between their compiled energies and our (any of the) calculations are up to 0.4 Ryd for some of the levels, such as 18--20. Therefore, there may be scope to improve upon our calculated energies but the (in)accuracy cannot definitely be determined by the limited comparison shown in Table~A. 

Our calculated energies from GRASP2 are listed in Table~1 along with those from FAC2 for the lowest 220 levels, which belong to the $n \le$ 3 configurations. Beyond these, the levels of the $n$ = 4 configurations start mixing. Discrepancies between the two sets of energies are smaller than   0.4 Ryd ($<$ 0.5\%) for a majority of  levels  and the orderings are different only in a few instances, such as  70/71 and 151/152. We also note that some differences may be because of a mismatch between the two sets of energies, as it is not always possible to perfectly match these due to their different notations. Also  note  that the $LSJ$ designations of the levels listed in Table~1 are not always unambiguous, and a few of these can be (inter)changed with varying amounts of CI, codes, and authors preferences. This is inevitable in any calculation because of the strong mixing among some of the levels. As examples, we list the lowest 20 levels in Table~B. For some, such as 1, 2, 10, and 12, there is a clear dominance of one vector (level) and hence there is no scope for ambiguity. However, for  others, such as 3--9, several vectors (levels) dominate and therefore it is not straightforward to designate such levels. For example, the eigenvector for level 19 is dominant in 19 but is also significant   in  4. However,  the eigenvector for level 105 is dominant in both levels 4 and 105 (not listed in Table~B). Finally, it may be noted that the degeneracy among the levels of W~LIX is very large -- see for example levels 3, 5, 9, 19, and 32 of 3s$^2$3p$^3$3d $^5$D$^o$, which are separated by up to $\sim$30 Ryd. For the ground state energy the Breit and QED contributions are 28.7 and 21.7 Ryd, respectively, although they amount to only $\sim$0.1\%.
 
\subsection{P-like W~LX}
 For this ion we have also performed two calculations with {\sc grasp} using different levels of CI, i.e. GRASP1:  includes 1313 levels of the 15 $n$ = 3 configurations, which are 3s$^2$3p$^3$, 3s$^2$3p$^2$3d, 3s3p$^4$, 3s$^2$3p3d$^2$, 3s3p$^3$3d, 3s3p$^2$3d$^2$, 3p$^5$, 3p$^4$3d, 3s$^2$3d$^3$, 3p$^3$3d$^2$, 3s3p3d$^3$, 3p$^2$3d$^3$, 3s3d$^4$, 3p3d$^4$, and 3d$^5$. In the other calculation (GRASP2), a further 20 configurations of [3s$^2$3p$^2$, 3s3p$^3$, 3s$^2$3p3d, 3s3p$^2$3d, and 3s$^2$3d$^2$]4$\ell$ are included, generating in total 3533 levels. Similarly, two calculations with {\sc fac} are performed, i.e. FAC1 with the same CI as in GRASP2, and FAC2, which also includes all possible combinations of 3$\ell^4$4$\ell$, generating 14~608 levels in total. Energies for the lowest 220 levels from both GRASP2 and FAC2 are listed in Table~2. These levels belong to the first 8 configurations listed above. For the higher-lying levels,   those of $n$ = 4 intermix with $n$ = 3. 
 
In Table~C we compare our energies for the lowest 25 levels of W~LX from GRASP1, GRASP2, FAC1, and FAC2  with the NIST compilation. CI for W~LX is  not as important as for W~LIX, because differences between our GRASP1 and GRASP2 energies are smaller than 0.02 Ryd. Similarly, discrepancies between the FAC1 and FAC2 energies are less than 0.03 Ryd. However, differences between the GRASP2 and FAC2 energies are up to 0.3 Ryd for some levels, for  reasons already explained in section 2.1. The NIST compilation is  only for the lowest 25 levels, listed in Table~C, and our GRASP2 energies are (generally) lower by up to 0.3 Ryd -- see for example, levels 13, 17 and 22. Similar differences remain between the NIST and FAC2 energies, and therefore are not due to a lack of CI. However, it is  worth emphasising  that the compiled energies of NIST are mostly based on  interpolation/extrapolation and hence are likely not very accurate. More importantly, there are  differences in the designations of a few levels, particularly the ground state, which is (3s$^2$3p$^3$)~$^2$D$^o_{3/2}$ in our work, but  $^2$P$^o_{3/2}$ in NIST. This is a highly mixed level and the eigenvector for $^2$P$^o_{3/2}$ dominates in both levels 1 and 25 -- see Table~D in which eigenvectors for the lowest 25  are listed. However, we have preferred to designate the lower (ground) level as $^2$D$^o_{3/2}$, because the placings of $^2$D$^o_{5/2}$ and $^2$P$^o_{1/2}$ (levels 5 and 6) are unambiguous. There may be  similar differences in designations with other calculations because of the very high mixing among some of the levels of W~LX.  
 
\subsection{Si-like W~LXI}
 
As for other W ions, we have performed two calculations each with the {\sc grasp} and {\sc fac} codes  to assess the effect of CI. These are GRASP1: 518 levels of 12 configurations [3s$^2$3p$^2$, 3s3p$^3$, 3s$^2$3p3d, 3s3p$^2$3d, 3p$^4$, 3s$^2$3d$^2$, 3p$^3$3d, 3s3p3d$^2$, 3p$^2$3d$^2$, 3s3d$^3$, 3p3d$^3$, and 3d$^4$]; GRASP2: 4364 levels of 48 configurations, the additional 36 are [3s$^2$3p, 3s3p$^2$, 3s$^2$3d, 3s3p3d, 3p$^3$, 3p$^2$3d,  3s3d$^2$, 3p3d$^2$,  and 3d$^3$]4$\ell$; FAC1: 9798 levels of 3*4, 3*3 4*1 and 3*4 5*1; and finally FAC2: which includes 27~122 levels in total, the additional ones arising from 3*3 6*1 and 3*2 4*2 configurations. Energies obtained from these calculations are compared in Table~E with the NIST compilation for the lowest 21 levels of W~LXI, which are the only ones in common. As for other ions, the CI is not very important for this ion, because the GRASP1 and GRASP2 energies agree within to 0.02 Ryd, and the FAC1 and FAC2 energies show no appreciable differences. Similarly, the agreement between our GRASP2 and FAC2 energies is better than 0.2 Ryd -- see levels 12--15. However, as for other ions, the differences with the NIST compilation are larger, up to 0.4 Ryd -- see level 9 for example. Again, the NIST energies are not very accurate and therefore such differences are not surprising. An important difference between our calculations and the NIST compilation is the designation for level 4, i.e. (3s3p$^3$)~$^5$S$^o_2$ which is $^3$P$^o_2$ (64) in the latter. Both these levels are highly mixed, as may be seen from the eigenvectors listed in Table~F for the lowest 21 levels {\em plus} the remaining two of the 3s3p$^3$ configuration, i.e. $^3$P$^o_2$ and $^1$P$^o_1$.
 
 Our recommended energies for the lowest 215 levels of W~LXI are listed in Table~3 from the GRASP2 and FAC2 calculations. These levels belong to the $n$ = 3 configurations and beyond these those of $n$ = 4 intermix. Finally, there are no major differences in the orderings of the two sets of level energies.
 
\subsection{Al-like W~LXII}
 
For W~LXII  the experimental energies are also as sparse as for other W ions. However, two sets of theoretical energy levels \cite{ajm,saf} are available in the literature. Safronova and Safronova \cite{saf} adopted a relativistic many-body perturbation theory (RMBPT) and reported energies for the lowest 40 levels belonging to the 3s$^2$3p, 3s3p$^2$, 3s$^2$3d,  3s3p3d, 3p$^3$, and 3p$^2$3d configurations.  In addition, S.~Aggarwal et al. \cite{ajm} have calculated  energies for the lowest 148 levels of the 3s$^2$3p, 3s3p$^2$, 3s$^2$3d, 3s3p3d, 3p$^3$, 3p$^2$3d, 3s3d$^2$, 3p3d$^2$, and 3d$^3$ (nine) configurations, adopting the same version of the {\sc grasp} code as in the present work. The RMBPT energies \cite{saf} are closer to the NIST compilation and in general are lower than those of S.~Aggarwal et al.  by up to 0.4 Ryd -- see Table~2 of \cite{ajm}. 

We have performed several sets of calculations with the {\sc grasp} code but mention only three here, namely: GRASP1, which includes the basic 148 levels of the 9 configurations listed above; GRASP2, which considers an  additional 776 (total 924) levels of the [3s3p, 3s3d, 3p3d, 3s$^2$, 3p$^2$, and 3d$^2$]4$\ell$ (24) configurations; and finally GRASP3 which includes a further 1079 levels (total 2003) of the 30 additional  configurations, i.e. [3s3p, 3s3d, 3p3d, 3s$^2$, 3p$^2$, and 3d$^2$]5$\ell$. S.~Aggarwal et al. \cite{ajm} included CI among  35 configurations, which are the basic 9 of GRASP1 {\em plus} another 26, i.e.  3s3p4$\ell$, 3s3d4$\ell$, 3p3d4$\ell$, 3s$^2$4$\ell$, 3p$^2$4$\ell$ (except 3p$^2$4d), 3p4$\ell^2$ (except 3p4p$^2$), and 3d4$\ell^2$. It is not clear  why they overlooked   configurations such as: 3p$^2$4d, 3p4p$^2$, 3s4$\ell^2$, and 3$\ell$4$\ell\ell'$. In addition, their  35 configurations generate 1007 levels in total (see Table~1 of \cite{km}) whereas they mention only 894, and therefore there is an anomaly of 113 levels. However, we  stress  that (particularly) the omission of the 3p$^2$4d and  3p4p$^2$ configurations does not affect the energies or the corresponding lifetimes, as already discussed by one of us \cite{km}. More importantly, levels of the 3$\ell$4$\ell^2$ configurations lie at energies well above those of our GRASP3 calculations, and hence are omitted from our work. This has been confirmed by our larger calculation with 75 configurations and 2393 levels. For the same reason we preferred not to include the 4$\ell^2$ configurations for the calculations of energy levels for other W ions. A complete set of energies for all 148 levels (of the GRASP1 calculations) are listed in Table~4 from GRASP3 and FAC2 (see below). We  note that levels from all other configurations clearly lie {\em above} these 148 and hence there is no intermixing.  

As with {\sc grasp}, we have also performed several calculations with {\sc fac}, but focus on only two, i.e. FAC1:  includes the same 2003 levels as in GRASP3, and FAC2: contains 12~139 levels in total, the additional ones arising from the 3*2 6*1, 3*1 4*2, 3*1 5*2 and 3*1 6*2 configurations. In Table~G we compare our energies from GRASP2, GRASP3, FAC1, and FAC2 with those of NIST for the lowest 21 levels, which are in common. Also included in this table are the results of Safronova and Safronova \cite{saf} from RMBPT. The corresponding data of S.~Aggarwal et al. \cite{ajm} are not considered  because they are similar to our GRASP2 calculations and have already been discussed previously \cite{km}. Although a considerably large CI has been included in our calculations, it does not appear to be too important for W~LXII, because the GRASP2 and GRASP3 (and FAC1 and FAC2) energies are practically identical. Therefore, the discrepancies  between the GRASP and FAC energies (up to 0.4 Ryd, particularly for level 21) are not due to different levels of CI but because of the computational and theoretical dissimilarities in the codes. Nevertheless, although the NIST energies are not claimed to be very accurate, their agreements with those from FAC and RMBPT are better (within 0.1 Ryd) than with GRASP. Regarding all the 148 levels in Table~4, the differences between the GRASP and FAC energies are up to 0.4  Ryd for some (see levels 77 upwards in the table).
  
Finally, as for other W ions, configuration mixing is strong for W~LXII also and therefore there is always a possibility of (inter)change of level designations listed in Table~4.  For the 21 levels listed in Table~G, their designations and orderings are the same between NIST and our calculations, but differ with those of S.~Aggarwal et al. \cite{ajm} for some, such as levels 10 and 68, i.e.  (3p$^3$) $^2$D$^o_{3/2}$ and   $^2$P$^o_{3/2}$, which are reversed by them. These two levels (and many more) have strong mixing, as may be seen from Table~H in which we list the eigenvectors for the lowest 21 levels plus 68, i.e. 3p$^3$~$^2$P$^o_{3/2}$. Similarly, there is a {\em disagreement} for most level designations between our work and NIST with those of  Safronova and Safronova \cite{saf}.
 
\subsection{Mg-like W~LXIII}
For this ion, earlier  calculations for energy levels are by Safronova and Safronova \cite{saf} using the RMBPT method for the lowest 35 levels of the 3s$^2$, 3s3p, 3p$^2$, 3s3d, 3p3d, and 3d$^2$ configurations, whereas the NIST compilation is only for 9 levels -- see Table~I. As for other ions we have performed several sets of calculations with {\sc grasp} and {\sc fac} and here we only state our final results. For the GRASP calculations we have considered 58 configurations, which are 3$\ell^2$, 3s3p, 3s3d, 3p3d, 3$\ell$4$\ell$, 4$\ell^2$, 4$\ell\ell'$, 3$\ell$5$\ell$, and 3$\ell$6$\ell$ (except 6h), while for  FAC  we  include 991 levels, the additional ones arising from 3$\ell$7$\ell$ and 4$\ell$5$\ell$. However, levels of the 4$\ell^2$, 4$\ell\ell'$ and 4$\ell$5$\ell$ configurations mostly lie above those of 3$\ell$7$\ell$ and can therefore be neglected. Energy levels from both calculations are listed in Table~5 for the lowest 210 levels. In Table~I a comparison is shown  for the lowest 35 levels with the NIST compilation and the RMBPT calculations \cite{saf}. As for W~LXII, the FAC and RMBPT energies agree closely with each other as well as with NIST, but our GRASP energies are higher by up to 0.3 Ryd for many levels. Similarly, mixing for the levels is strong for a few as shown in Table~J for the lowest 35 -- see in particular levels 22, 25 and 34.

\subsection{Na-like W~LXIV}
For this ion we have gradually increased the number of orbitals to perform {\sc grasp} calculations for up to 1235 levels. The configurations included are 2p$^6$$n\ell$ with $n \le$ 7 and $\ell \le$ 4, 2p$^5$3$\ell\ell'$, 2p$^5$3$\ell^2$, 2p$^5$4$\ell\ell'$, 2p$^5$4$\ell^2$, and 2p$^5$3$\ell$4$\ell$. However, we note that the levels of 2p$^6$$n\ell$ lie {\em below} those of the other configurations. For this reason we only list the lowest 30 levels in Table~K, all belonging to  2p$^6$$n\ell$. However, with {\sc fac} we have performed comparatively larger calculations for up to $n$ = 20 and all possible values of $\ell$, i.e. 1592 levels in total. These results are also listed in Table~K along with those of NIST, which are confined to the $n \le$ 5 levels.  The NIST  energies differ with FAC by up to 0.26 Ryd for some levels (see 20), but discrepancies are smaller than  0.15 Ryd with those with {\sc grasp}. Again, the differences between the GRASP and FAC energies are not because of different levels of CI, but due to methodological variations. It has not been possible to include higher 2p$^6$$n\ell$ configurations in our {\sc grasp} calculations, but since the {\sc fac}  energies have been obtained  (as stated above) in Table~6 we list  these for the lowest 396 levels, all belonging to 2p$^6$$n\ell$ with $n \le$ 20. This will be helpful for future comparisons. Finally, unlike the other W ions discussed above, there is no (strong)  mixing and/or ambiguity for the designation of the 2p$^6$$n\ell$ levels listed in Tables~K and 6. 

Safronova et al. \cite{saf2} have reported energies for 242 levels of W~LXIV from three independent codes, namely RMBPT, HULLAC (Hebrew University Lawrence Livermore Atomic Code \cite{hullac}) and the atomic structure code of R.D.~Cowan  available at {\tt  http://das101.isan.troitsk.ru/cowan.htm}. Although NIST energies for this ion are only available for a few levels, as already seen in Table~K, their RMBPT results are  closest to the measurements. Additionally, based on the comparisons made for other W ions, their RMBPT energies should be the most accurate. Nevertheless, the RMBPT energy for level 2 (2p$^5$3s $^3$P$^o_2$) differs by 1.3\% and 6.4\% with those from HULLAC and Cowan, respectively. Corresponding differences for the remaining levels are up to 0.3\% and 1\%, respectively. Only the lowest 5 levels of Table~K are common with their work, as the remaining 237 belong to the 2p$^5$3$\ell\ell'$ configurations. Therefore, our listed energies in Table~6 supplement their data.

\subsection{Ne-like W~LXV}
The NIST compilation of energies for this ion is limited to only 10 levels of the 2p$^5$3$\ell$ configurations. However, Vilkas et al. \cite{mrmp} have reported energies for 141 levels of the 2p$^6$,  (2s2p$^6$)3$\ell$, 4$\ell$, 5$\ell$ (except 5g), and  (2p$^5$) 3$\ell$, 4$\ell$, 5$\ell$ (except 5g) configurations. For their calculations they adopted the relativistic multi-reference many-body M{\o}ller-Plesset (MRMP) perturbation theory, and included CI up to the $n$ = 5 orbitals. We have included the same configurations for our calculations with {\sc grasp}, which generate 157 levels in total because we have also considered the 5g orbital. However, in Table~7 we list energies for only the lowest 121, because beyond this the levels of  the 2s2p$^6$6$\ell$ configurations start mixing in the same way as of 2s2p$^6$5g with those of 2s2p$^6$4$\ell$ -- see levels 92--99 in the table. Additionally, we have performed larger calculations with {\sc fac} with up to 1147 levels, belonging to the 2*8, (2*7) 3*1, 4*1, 5*1, 6*1, 7*1, and 2*6 3*2 configurations. These results are also listed in Table~7 for  comparison. Differences between the GRASP and FAC energies are up to 0.5 Ryd (0.07\%) for some levels, but the level orderings are almost identical. Similarly, there is no difference in level orderings with the MRMP calculations \cite{mrmp} and the energies differ only by less than 0.6 Ryd (0.06\%) with GRASP -- see levels 63 and 77--83. Therefore, overall there is no (significant) discrepancy between the three independent calculations. However, in general the FAC energies are lower than those from GRASP for a majority of levels, whereas those of MRMP are higher.

In Table~L, we compare energies with the NIST compilation for only the {\em common} levels. There is no uniform pattern for (dis)agreement between the theoretical and experimental energies. In general, the MRMP energies are closer to those of NIST whereas those from FAC differ the most. Unfortunately, these comparisons are not sufficient for  accuracy determination, particularly when the NIST energies are not based on direct measurements. Finally, as for most W ions, for W~LXV also there is a strong mixing for some levels and therefore the level designations listed in Table~7 can vary, although the MRMP  calculations \cite{mrmp} have the same labels as in our work. Nevertheless, in Table~M we list the eigenvectors for the lowest 33 levels, which include all of the NIST compilation. Note particularly the mixing for levels 24, 25 and 31.

\subsection{F-like W~LXVI}
 For this ion we have performed a series of calculations with {\sc grasp} with gradually increasing  CI and our final set includes 501 levels of 38 configurations, which are: 2s$^2$2p$^5$, 2s2p$^6$, (2s$^2$2p$^4$, 2s2p$^5$, 2p$^6$)3$\ell$, 4$\ell$, 5$\ell$. Similarly, calculations with {\sc fac} have been performed for up to 1113 levels from the 2*7 and (2*6) 3*1, 4*1, 5*1, 6*1, 7*1 configurations. These levels span an energy range of up to 1360 Ryd. Opening the 1s shell gives rise to levels above 5000 Ryd and therefore has not been included in the calculations. Energies from both of these calculations are listed in Table~8 for the lowest 150 levels, because beyond this the levels  of the $n$ = 5 configurations start mixing. However, the listed levels include all of the $n$ = 3 configurations. Differences between the two sets of energies are up to 0.5 Ryd for some levels, except three (145--147) for which the discrepancies are slightly larger, up to 0.7 Ryd. The level orderings are also the same for a majority of levels, but slightly differ in a few instances, such as for 93--112. NIST listings are available for only two levels, namely 2s$^2$2p$^5$~$^2$P$^o_{1/2}$ and 2s2p$^6$~$^2$S$_{1/2}$, and the energy for the latter is lower by 0.5 Ryd than the theoretical results. No other similar theoretical energies are available for this ion for comparison purposes. Finally, this ion is no exception for level mixing and examples of this are listed in Table~N for the lowest 48 levels --  see in particular   13, 15, 40, and 42.
 
\section{Radiative rates}\label{sec.eqs}
Apart from energy levels, calculations have  been made for absorption oscillator strengths ($f$-values, dimensionless), radiative rates ($A$-values, s$^{-1}$) and line strengths ($S$-values, in atomic units, 1 a.u. = 6.460$\times$10$^{-36}$ cm$^2$ esu$^2$). However, $f$- and $A$-values  for all types of  transition ($i \to j$)  are connected by the following expression:

\begin{equation}
f_{ij} = \frac{mc}{8{\pi}^2{e^2}}{\lambda^2_{ji}} \frac{{\omega}_j}{{\omega}_i}A_{ji}
 = 1.49 \times 10^{-16} \lambda^2_{ji} \frac{{\omega}_j}{{\omega}_i} A_{ji}
\end{equation}
where $m$ and $e$ are the electron mass and charge, respectively, $c$  the velocity of light,  $\lambda_{ji}$  the transition wavelength in $\rm \AA$, and $\omega_i$ and $\omega_j$  the statistical weights of the lower $i$ and upper $j$ levels, respectively. Similarly, $f$- and  $A$-values are related to $S$ by the  standard equations given in \cite{w40b}. 

In Tables 9--16 we present results  for energies (wavelengths, $\lambda_{ji}$ in ${\rm \AA}$), $A$-, $f$- and $S$- values for electric dipole (E1) transitions in W ions, which have been obtained with the {\sc grasp} code. For other types of transitions, namely  magnetic dipole (M1), electric quadrupole (E2), and magnetic quadrupole (M2), only the $A$-values are listed, because the corresponding results for $f$- or $S$-values can be  obtained using Eqs. (1-5) given in \cite{w40b}. Additionally, we have also listed the ratio (R) of the  velocity (Coulomb gauge) and length (Babushkin gauge) forms which often (but not necessarily) give an indication of the accuracy. The {\em indices} used to represent the lower and upper levels of a transition are defined in Tables 1--8. Furthermore, only a limited range of transitions are listed in Tables 9--16, but full tables are available online in the electronic version.

For the W ions considered here, existing $A$- (or $f$-) values are available mostly for three ions, i.e. Al-like W~LXII \cite{saf}, Mg-like W~LXIII  \cite{saf} and Na-like  W~LXIV \cite{mrmp}. Therefore, we confine our comparisons to these three ions. In Table~O we compare the $f$-values for common E1 transitions with the results of Safronova and Safronova \cite{saf}. Both sets of data agree very well for all transitions.  Similarly, for a few weak transitions ($f$ $\sim$ 10$^{-4}$), such as 1--22, 2--3  and 14--19, the ratio R is up to 1.7 and is closer to unity for the comparatively strong transitions. Similar comparison with their results for transitions in W~LXIII is shown in Table~P. For the common transitions listed here, R is unity for all, and $f$-values agree closely for most with only a few exceptions, such as 20--32, 21--30 and  26--34 for which discrepancies  are  a factor of two. However, we note that the $f$- (or $A$-) values of \cite{saf} are only for a small number of transitions whereas our results listed in Tables 12 and 13 cover a much wider range.    

Vilkas et al. \cite{mrmp} have listed $A$-values for some (not all) transitions of W~LXV and in Table~Q we compare their results with our calculations with {\sc grasp}, but only from the lowest three to higher excited levels. Additionally we have  listed the $f$-values to indicate the strength of transitions. As for other W ions, R is also listed for these transitions  and is within a few percent of unity, irrespective of the $f$-value. There are no appreciable differences between the two sets of $A$-values and discrepancies, if any, are (generally) within $\sim$20\%.

The comparisons of $A$- ($f$-) values discussed above are only for a subset of transitions. Considering a wider range, for a majority of  strong transitions ($f$ $\ge$ 0.01) R is often within 20\% of unity, as already seen in Tables~O, P and Q. However, there are (as always) some exceptions. For example, there are only six transitions of W~LXIII  with  $f$ $>$ 0.01 for which R is up to 1.6, namely 148--166 ($f$ = 0.011, R = 1.3),  158--173 ($f$ = 0.021, R = 1.3),  160--174  ($f$ = 0.028, R = 1.6),  161--175  ($f$ = 0.025, R = 1.4),  162--176  ($f$ = 0.027, R = 1.4),  and 163--177  ($f$ = 0.029, R = 1.6). Therefore, based on this and other comparisons already discussed, our assessment of accuracy for the $f$-values for a majority of strong transitions is $\sim$20\%. Finally, for much weaker transitions  (often with $f$ $\le$ 10$^{-4}$), R can be several orders of magnitude and it is very difficult to assess the accuracy of the $f$-values because results are often much more variable with CI and/or codes. Generally, such transitions do not make an appreciable contribution to plasma modelling and their results are mostly required for completeness. 

\section{Lifetimes}

The lifetime $\tau$ of a level $j$ is given by 1.0/$\Sigma_{i}$$A_{ji}$ and the summation includes $A$-values from all types of transitions, i.e. E1, E2, M1, and M2. Since this is a measurable quantity it helps to assess the accuracy of $A$-values, particularly when a single (type of) transition dominates. Unfortunately, to our knowledge no measurements of $\tau$ are available for the levels of the W ions considered here, but in Tables 1--8  we   list our calculated results.  Previous theoretical results are available for two ions, i.e. W~LXII \cite{ajm} and W~LXV \cite{mrmp}. Unfortunately, the $\tau$ of S.~Aggarwal et al. \cite{ajm} contain large errors, by up to 14 orders of magnitude,  for over 90\%  of the levels of W~LXII and bear no relationship to the $A$-values, as already discussed \cite{km}. For W~LXV, the reported $\tau$ of Vilkas et al. \cite{mrmp} are included in Table~7, and there is no significant discrepancy for any level.

\section{Conclusions}

Energy levels and radiative rates for E1, E2, M1, and M2 transitions are reported  for  eight W ions (W~LIX to W~LXVI). A large number of levels are considered for each ion and the data sets reported here are significantly larger than available in the literature.  For our calculations the {\sc grasp}  code has been adopted, although {\sc fac} has also been utilised for the determination of energy levels to assess the importance of CI, larger than that considered in  {\sc grasp}. It is concluded that CI beyond a certain level does not appreciably improve the level energies. Differences between the GRASP and FAC energies, and the available  experimental and theoretical values, are often smaller than 0.5 Ryd, or equivalently the listed energy levels for all W ions are assessed to be accurate to better than 1\%, but   scope remains for improvement.  A similar assessment of accuracy for the corresponding $A$-values is not feasible, mainly because of the paucity of other comparable  results. However, for strong transitions (with large $f$-values),   the accuracy for $A$-values and  lifetimes may be $\sim$20\%. 
  Lifetimes for these levels are also listed although no measurements are currently available in the literature. However, previous theoretical values are available for most levels of W~LXV and there is no discrepancy with our work.



\ack
KMA  is thankful to  AWE Aldermaston for financial support. 

\begin{appendix}

\def\thesection{} 

\section{Appendix A. Supplementary data}

Owing to space limitations, only parts of Tables 9--16  are presented here, the full tables being made available as supplemental material in conjunction with the electronic
publication of this work. Supplementary data associated with this article can be found, in the online version, at doi:nn.nnnn/j.adt.2016.nn.nnn.

\end{appendix}



\clearpage
\newpage


\renewcommand{\baselinestretch}{1.0}
\footnotesize
\begin{longtable}{@{\extracolsep\fill}rlllrrrr@{}}
\caption{Comparison of threshold energies (in Ryd) for the lowest 20 levels of  W LIX.} 
Index  &     Configuration                    & Level            & NIST      & GRASP1   &   GRASP2 & FAC1      & FAC2  \\  \\
 \hline\\
\endfirsthead\\
\caption[]{(continued)}
Index  &     Configuration                    & Level            & NIST      & GRASP1   &   GRASP2 & FAC1      & FAC2  \\  \\
\hline\\
\endhead 
    1   &  3s$^2$3p$^4$ 		      &   $^3$P$  _{2}$  &   00.0000 &  0.0000  &   0.0000 &   0.00000  &  0.0000    \\
    2   &  3s$^2$3p$^4$ 		      &   $^1$S$  _{0}$  &   01.394  &  1.4629  &   1.4691 &   1.46006  &  1.4675    \\
    3   &  3s$^2$3p$^3$($^4$S)3d	      &   $^5$D$^o_{2}$  &   17.2585 & 17.3063  &  17.2920 &  17.23239  & 17.2200    \\ 
    4   &  3s$^2$3p$^3$($^2$D)3d	      &   $^3$P$^o_{0}$  &   17.852  & 17.9536  &  17.9413 &  17.85940  & 17.8689    \\
    5   &  3s$^2$3p$^3$($^4$S)3d	      &   $^5$D$^o_{1}$  &   17.852  & 17.9342  &  17.9213 &  17.87821  & 17.8489    \\
    6   &  3s$^2$3p$^3$($^2$P)3d	      &   $^3$F$^o_{3}$  &   17.9173 & 17.9812  &  17.9637 &  17.90613  & 17.8913    \\
    7   &  3s$^2$3p$^3$($^2$D)3d	      &   $^1$G$^o_{4}$  &   23.4589 & 23.4945  &  23.4758 &  23.42838  & 23.4095    \\
    8   &  3s$^2$3p$^3$($^2$D)3d	      &   $^3$D$^o_{2}$  &   23.94   & 23.9785  &  23.9637 &  23.91315  & 23.8975    \\
\hline            								                	 
\end{longtable}   								   					       
			      							   					       

														       
\begin{flushleft}													       
{\small
NIST:  {\tt http://www.nist.gov/pml/data/asd.cfm} \\
GRASP1: Present results with the {\sc grasp} code from 18  configurations and 2762 levels \\				      		      
GRASP2: Present results with the {\sc grasp} code from 46  configurations and 12~652 levels \\
FAC1: Present results with the {\sc fac} code  from  2762 levels \\
FAC2: Present results with the {\sc fac} code  from  38~694 levels \\		       
}															       
\end{flushleft} 


\renewcommand{\baselinestretch}{1.0}
\footnotesize
\begin{longtable}{@{\extracolsep\fill}rllr@{}}
\caption{Eigenvectors (EV) for the lowest 20 levels of W~LIX from the {\sc grasp} code. Numbers outside and inside a bracket correspond to EV and the level, \\respectively. See Table~1 for the definition of all levels.}
Index  & Configuration           & Level              &  Eigenvectors     \\  \\
 \hline\\
\endfirsthead\\
\caption[]{(continued)}
Index  & Configuration           & Level              &  Eigenvectors   \\  \\
\hline\\
\endhead
    1   &  3s$^2$3p$^4$ 		      &   $^3$P$  _{2}$  &      0.67(  1)+0.31( 12)									   \\
    2   &  3s$^2$3p$^4$ 		      &   $^1$S$  _{0}$  &      0.36( 46)+0.64(  2)									   \\
    3   &  3s$^2$3p$^3$($^4$S)3d	      &   $^5$D$^o_{2}$  &      0.16(  3)+0.08( 41)+0.19( 18)+0.06( 45)+0.10( 26)+0.23(108)+0.12(132)			   \\
    4   &  3s$^2$3p$^3$($^2$D)3d	      &   $^3$P$^o_{0}$  &      0.26( 19)+0.14(  4)+0.12( 37)+0.48(105) 						   \\
    5   &  3s$^2$3p$^3$($^4$S)3d	      &   $^5$D$^o_{1}$  &      0.24(  5)+0.14( 20)+0.10( 43)+0.18( 30)+0.27(107)					   \\
    6   &  3s$^2$3p$^3$($^2$P)3d	      &   $^3$F$^o_{3}$  &      0.13(  9)+0.12(106)+0.21( 22)+0.05( 44)+0.27(  6)+0.06(136)+0.14( 40)			   \\
    7   &  3s$^2$3p$^3$($^2$D)3d	      &   $^1$G$^o_{4}$  &      0.26( 32)+0.14( 25)+0.10(  7)+0.48(130) 						   \\
    8   &  3s$^2$3p$^3$($^2$D)3d	      &   $^3$D$^o_{2}$  &      0.10(  3)+0.14( 41)+0.16(  8)+0.07( 28)+0.04(108)+0.22( 39)+0.19(132)			   \\
\hline  											       								   					      
\end{longtable}

                               
\begin{flushleft}                                                                               
                               
{\small
                                                                               
}                                                                                               
                               
\end{flushleft} 

\renewcommand{\baselinestretch}{1.0}
\footnotesize
\begin{longtable}{@{\extracolsep\fill}rlllrrrr@{}}
\caption{Comparison of threshold energies (in Ryd) for the lowest 25 levels of  W LX.}
Index  &     Configuration          & Level              & NIST  & GRASP1   & GRASP2 & FAC1     &   FAC2  \\  \\
 \hline\\
\endfirsthead\\
\caption[]{(continued)}
Index  &     Configuration          & Level              & NIST  & GRASP1   & GRASP2 & FAC1     &   FAC2  \\  \\
\hline\\
\endhead 
    1   & 3s$^2$3p$^3$  	   &   $^2$D$^o_{3/2 }$  &  00.0000 &  0.0000  &     0.0000  &  0.0000  &     0.0000   \\
    2   & 3s$^2$3p$^2$($^3$P)3d    &   $^4$F$  _{3/2 }$  & 16.8821  & 16.9403  &    16.9357  & 16.8671  &    16.8529   \\
    3   & 3s$^2$3p$^2$($^1$S)3d    &   $^2$D$  _{5/2 }$  & 23.903   & 23.8688  &    23.8613  & 23.8040  &    23.7839   \\
    4   & 3s$^2$3p$^3$  	   &   $^4$S$^o_{3/2 }$  & 25.060   & 25.0648  &    25.0619  & 25.0905  &    25.0865   \\
    5   & 3s$^2$3p$^3$  	   &   $^2$D$^o_{5/2 }$  & 25.9556  & 25.9490  &    25.9411  & 25.9727  &    25.9639   \\
    6   & 3s$^2$3p$^3$  	   &   $^2$P$^o_{1/2 }$  & 27.019   & 27.0845  &    27.0831  & 27.1065  &    27.1049   \\
    7   & 3s3p$^4$		   &   $^4$P$  _{5/2 }$  & 37.9315  & 37.9970  &    38.0129  & 37.9087  &    37.9134   \\
    8   & 3s3p$^4$		   &   $^2$P$  _{3/2 }$  & 40.242   & 40.1880  &    40.1973  & 40.1132  &    40.1082   \\
\hline            								                	 
\end{longtable}   								   					       
			      							   					       

														       
\begin{flushleft}													       
{\small
NIST:  {\tt http://www.nist.gov/pml/data/asd.cfm} \\
GRASP1: Present results with the {\sc grasp} code from 15  configurations and 1313 levels \\
GRASP2: Present results with the {\sc grasp} code from 35  configurations and 3533 levels \\
FAC1: Present results with the {\sc fac} code  from  1313 levels \\
FAC2: Present results with the {\sc fac} code  from  14~608 levels \\		       
}															       
\end{flushleft} 

\clearpage
\newpage

\renewcommand{\baselinestretch}{1.0}
\footnotesize
\begin{longtable}{@{\extracolsep\fill}rllr@{}}
\caption{Eigenvectors (EV) for the lowest 25 levels of W~LX from the {\sc grasp} code. Numbers outside and inside a bracket correspond to EV and the level, \\respectively. See Table~2 for the definition of all levels.}
Index  & Configuration           & Level              &  Eigenvectors    \\ \\
 \hline\\
\endfirsthead\\
\caption[]{(continued)}
Index  & Configuration           & Level              &  Eigenvectors   \\ \\
\hline\\
\endhead
     1   & 3s$^2$3p$^3$  	   &   $^2$D$^o_{3/2 }$  &   	  0.25(  4)+0.27(  1)+0.48( 25) 					\\
    2   & 3s$^2$3p$^2$($^3$P)3d    &   $^4$F$  _{3/2 }$  &   	  0.34(  2)+0.12( 12)+0.10( 23)+0.11( 18)+0.31( 60)			\\
    3   & 3s$^2$3p$^2$($^1$S)3d    &   $^2$D$  _{5/2 }$  &   	  0.17( 10)+0.20( 57)+0.14( 21)+0.15( 20)+0.30(  3)			\\
    4   & 3s$^2$3p$^3$  	   &   $^4$S$^o_{3/2 }$  &   	  0.55(  4)+0.45(  1)							\\
    5   & 3s$^2$3p$^3$  	   &   $^2$D$^o_{5/2 }$  &   	  1.00(  5)								\\
    6   & 3s$^2$3p$^3$  	   &   $^2$P$^o_{1/2 }$  &   	  0.98(  6)								\\
    7   & 3s3p$^4$		   &   $^4$P$  _{5/2 }$  &   	  0.66(  7)+0.27( 45)							\\
    8   & 3s3p$^4$		   &   $^2$P$  _{3/2 }$  &   	  0.08(  2)+0.06( 58)+0.11( 18)+0.06( 16)+0.11( 41)+0.32(  8)+0.24( 51) \\
\hline  											       								   					      
\end{longtable}

                               
\begin{flushleft}                                                                               
                               
{\small
                                                                               
}                                                                                               
                               
\end{flushleft} 

\clearpage
\newpage

\renewcommand{\baselinestretch}{1.0}
\footnotesize
\begin{longtable}{@{\extracolsep\fill}rlllrrrr@{}}
\caption{Comparison of threshold energies (in Ryd) for the lowest 21 levels of  W LXI.}
Index  &     Configuration          & Level              & NIST     & GRASP1   & GRASP2   & FAC1     &   FAC2  \\  \\
 \hline\\
\endfirsthead\\
\caption[]{(continued)}
Index  &     Configuration          & Level              & NIST     & GRASP1   & GRASP2   & FAC1     &   FAC2  \\  \\
\hline\\
\endhead 
    1   & 3s$^2$3p$^2$  	   &   $^3$P$  _{0}$  & 00.0000  & 00.0000  &  0.0000  &  0.0000  &   0.0000  \\
    2   & 3s$^2$3p$^2$  	   &   $^3$P$  _{1}$  & 25.5337  & 25.5392  & 25.5395  & 25.5650  &  25.5675  \\
    3   & 3s$^2$3p$^2$  	   &   $^1$D$  _{2}$  & 26.2946  & 26.2402  & 26.2351  & 26.2574  &  26.2587  \\
    4   & 3s3p$^3$		   &   $^5$S$^o_{2}$  & 38.2094  & 38.0581  & 38.0606  & 37.9699  &  37.9709  \\
    5   & 3s3p$^3$		   &   $^3$D$^o_{1}$  & 39.9800  & 40.1129  & 40.1071  & 40.0211  &  40.0210  \\
    6   & 3s$^2$3p3d		   &   $^3$F$^o_{2}$  & 	 & 41.7903  & 41.7744  & 41.7215  &  41.7222  \\
    7   & 3s$^2$3p3d		   &   $^3$D$^o_{1}$  & 	 & 45.1086  & 45.0826  & 45.0158  &  45.0129  \\
    8   & 3s$^2$3p3d		   &   $^3$P$^o_{2}$  & 49.57	 & 49.1925  & 49.1738  & 49.1246  &  49.1247  \\
\hline            								                	 
\end{longtable}   								   					       
			      							   					       

														       
\begin{flushleft}													       
{\small
NIST:  {\tt http://www.nist.gov/pml/data/asd.cfm} \\
GRASP1: Present results with the {\sc grasp} code from 12  configurations and 518 levels \\
GRASP2: Present results with the {\sc grasp} code from 48  configurations and 4364 levels \\
FAC1: Present results with the {\sc fac} code  from  9798 levels \\
FAC2: Present results with the {\sc fac} code  from  27~122 levels \\			      		      	       
}															       
\end{flushleft} 

\renewcommand{\baselinestretch}{1.0}
\footnotesize
\begin{longtable}{@{\extracolsep\fill}rllr@{}}
\caption{Eigenvectors (EV) for the lowest 21 (plus $^3$P$^o_2$ and $^1$P$^o_1$ of 3s3p$^3$) levels of W~LXI from the {\sc grasp} code. Numbers outside and inside a bracket \\correspond to EV and the level, respectively. See Table~3 for the definition of all levels.}
Index  & Configuration           & Level              &  Eigenvectors     \\ \\
 \hline\\
\endfirsthead\\
\caption[]{(continued)}
Index  & Configuration           & Level              &  Eigenvectors   \\  \\
\hline\\
\endhead
    1   & 3s$^2$3p$^2$  	   &   $^3$P$  _{0}$  &   0.69(  1)+0.31( 11)								  \\
    2   & 3s$^2$3p$^2$  	   &   $^3$P$  _{1}$  &   1.00(  2)									  \\
    3   & 3s$^2$3p$^2$  	   &   $^1$D$  _{2}$  &   0.35( 10)+0.64(  3)								  \\
    4   & 3s3p$^3$		   &   $^5$S$^o_{2}$  &   0.27(  4)+0.16( 16)+0.08( 20)+0.46( 64)					  \\
    5   & 3s3p$^3$		   &   $^3$D$^o_{1}$  &   0.14( 21)+0.28(  5)+0.16( 19)+0.24( 69)+0.09(  7)+0.05( 29)			  \\
    6   & 3s$^2$3p3d		   &   $^3$F$^o_{2}$  &   0.74(  6)+0.19( 27)								  \\
    7   & 3s$^2$3p3d		   &   $^3$D$^o_{1}$  &   0.04( 21)+0.04(  5)+0.05( 69)+0.44(  7)+0.14( 24)+0.25( 29)			  \\
    8   & 3s$^2$3p3d		   &   $^3$P$^o_{2}$  &   0.34( 23)+0.45(  8)+0.14( 27) 						  \\
\hline  											       								   					      
\end{longtable}

                               
\begin{flushleft}                                                                               
                               
{\small
                                                                               
}                                                                                               
                               
\end{flushleft} 

\renewcommand{\baselinestretch}{1.0}
\footnotesize
\begin{longtable}{@{\extracolsep\fill}rlllrrrrr@{}}
\caption{Comparison of threshold energies (in Ryd) for the lowest 21 levels of  W LXII.}
Index  &     Configuration          & Level           & NIST     & GRASP2   & GRASP3    & FAC1     &   FAC2   & RMBPT  \\  \\
 \hline\\
\endfirsthead\\
\caption[]{(continued)}
Index  &     Configuration          & Level           & NIST     & GRASP2   & GRASP3    & FAC1     &   FAC2   & RMBPT  \\  \\
\hline\\
\endhead 
    1   & 3s$^2$3p	      &     $^2$P$^o_{1/2 }$  & 00.0000  &  0.0000  &	0.0000  &  0.0000  &  0.0000  &   0.0000  \\
    2   & 3s3p$^2$	      &     $^4$P$  _{1/2 }$  & 12.3076  & 12.4425  &  12.4422  & 12.3220  & 12.3204  &  12.3018  \\
    3   & 3s$^2$3p	      &     $^2$P$^o_{3/2 }$  & 26.7311  & 26.7061  &  26.7060  & 26.7306  & 26.7314  &  26.7056  \\
    4   & 3s3p$^2$	      &     $^4$P$  _{3/2 }$  & 36.7742  & 36.8823  &  36.8835  & 36.7935  & 36.7959  &  36.7790  \\
    5   & 3s3p$^2$	      &     $^2$D$  _{5/2 }$  & 38.109   & 38.2380  &  38.2375  & 38.1447  & 38.1457  &  38.1110  \\
    6   & 3s3p$^2$	      &     $^2$D$  _{3/2 }$  & 39.6875  & 39.7869  &  39.7857  & 39.7028  & 39.7033  &  39.6637  \\
    7   & 3s3p$^2$	      &     $^2$P$  _{1/2 }$  & 40.4238  & 40.5824  &  40.5806  & 40.4826  & 40.4819  &  40.4024  \\
    8   & 3s$^2$3d	      &     $^2$D$  _{3/2 }$  & 43.9039  & 44.0171  &  44.0103  & 43.9482  & 43.9448  &  43.8775  \\
\hline            								                	 
\end{longtable}   								   					       
			      							   					       

														       
\begin{flushleft}													       
{\small
NIST:  {\tt http://www.nist.gov/pml/data/asd.cfm} \\
GRASP2: Present results with the {\sc grasp} code from 33  configurations and 928 levels \\
GRASP3: Present results with the {\sc grasp} code from 63  configurations and 2003 levels \\
FAC1: Present results with the {\sc fac} code  from  2003 levels \\
FAC2: Present results with the {\sc fac} code  from  	\\
RMBPT: Earlier results of Safronova and Safronova \cite{saf} \\ 	      		      	       
}															       
\end{flushleft} 

\clearpage
\newpage

\renewcommand{\baselinestretch}{1.0}
\footnotesize
\begin{longtable}{@{\extracolsep\fill}rllr@{}}
\caption{Eigenvectors (EV) for the lowest 21 (plus 68) levels of W~LXII from the {\sc grasp} code. Numbers outside and inside a bracket correspond to EV \\and the level, respectively. See Table~4 for the definition of all levels.}
Index  & Configuration           & Level              &  Eigenvectors     \\  \\
 \hline\\
\endfirsthead\\
\caption[]{(continued)}
Index  & Configuration           & Level              &  Eigenvectors  \\  \\
\hline\\
\endhead
    1   & 3s$^2$3p	      &     $^2$P$^o_{1/2 }$  &  1.00(  1)							       \\
    2   & 3s3p$^2$	      &     $^4$P$  _{1/2 }$  &  0.52(  2)+0.17(  7)+0.30( 20)  				       \\
    3   & 3s$^2$3p	      &     $^2$P$^o_{3/2 }$  &  0.98(  3)							       \\
    4   & 3s3p$^2$	      &     $^4$P$  _{3/2 }$  &  0.88(  4)+0.08(  6)						       \\
    5   & 3s3p$^2$	      &     $^2$D$  _{5/2 }$  &  0.37( 19)+0.61(  5)						       \\
    6   & 3s3p$^2$	      &     $^2$D$  _{3/2 }$  &  0.28( 22)+0.49(  6)+0.18(  8)  				       \\
    7   & 3s3p$^2$	      &     $^2$P$  _{1/2 }$  &  0.29(  2)+0.69(  7)						       \\
    8   & 3s$^2$3d	      &     $^2$D$  _{3/2 }$  &  0.08( 22)+0.12(  6)+0.79(  8)  				       \\
\hline     
\end{longtable}

                               
\begin{flushleft}                                                                               
                               
{\small
                                                                               
}                                                                                               
                               
\end{flushleft} 

\clearpage
\newpage

\renewcommand{\baselinestretch}{1.0}
\footnotesize
\begin{longtable}{@{\extracolsep\fill}rlllrrr@{}}
\caption{Comparison of threshold energies (in Ryd) for the lowest 35 levels of  W LXIII.}
Index  &     Configuration          & Level & NIST     & GRASP     & FAC        & RMBPT  \\  \\
 \hline\\
\endfirsthead\\
\caption[]{(continued)}
Index  &     Configuration          & Level & NIST     & GRASP     & FAC        & RMBPT  \\  \\
\hline\\
\endhead 
    1  &  3s$^2$       &    $^1$S$  _{0}$   & 00.0000  &   0.0000  &   0.0000	&   0.0000 \\
    2  &  3s3p         &    $^3$P$^o_{0}$   & 10.261   &  10.3595  &  10.2414	&  10.2650 \\
    3  &  3s3p         &    $^3$P$^o_{1}$   & 11.4036  &  11.5247  &  11.4028	&  11.4104 \\
    4  &  3p$^2$       &    $^3$P$  _{0}$   & 	       &  24.7520  &  24.5032	&  24.4911 \\
    5  &  3s3p         &    $^3$P$^o_{2}$   & 37.398   &  37.4521  &  37.3609	&  37.3992 \\
    6  &  3s3p         &    $^1$P$^o_{1}$   & 40.0821  &  40.2273  &  40.1296	&  40.1225 \\
    7  &  3p$^2$       &    $^1$D$  _{2}$   & 	       &  50.6187  &  50.4140	&  50.4418 \\
    8  &  3p$^2$       &    $^3$P$  _{1}$   & 	       &  50.7934  &  50.5757	&  50.5885 \\
\hline            								                	 
\end{longtable}   								   					       
			      							   					       

														       
\begin{flushleft}													       
{\small
NIST:  {\tt http://www.nist.gov/pml/data/asd.cfm} \\
GRASP: Present results with the {\sc grasp} code from 58  configurations and 509 levels \\
FAC: Present results with the {\sc fac} code  from  991 levels \\
RMBPT: Earlier results of Safronova and Safronova \cite{saf} \\ 		      		      	       
}															       
\end{flushleft} 

\clearpage
\newpage

\renewcommand{\baselinestretch}{1.0}
\footnotesize
\begin{longtable}{@{\extracolsep\fill}rllr@{}}
\caption{Eigenvectors (EV) for the lowest 35 levels of W~LXIII from the {\sc grasp} code. Numbers outside and inside a bracket correspond to EV \\and the level, respectively. See Table~5 for the definition of all levels.}
Index  & Configuration           & Level              &  Eigenvectors     \\  \\
 \hline\\
\endfirsthead\\
\caption[]{(continued)}
Index  & Configuration           & Level              &  Eigenvectors   \\  \\
\hline\\
\endhead
    1  &  3s$^2$       &    $^1$S$  _{0}$   &     1.00(  1)				   \\
    2  &  3s3p         &    $^3$P$^o_{0}$   &     1.00(  2)				   \\
    3  &  3s3p         &    $^3$P$^o_{1}$   &     0.72(  3)+0.27(  6)			   \\
    4  &  3p$^2$       &    $^3$P$  _{0}$   &     0.69(  4)+0.30( 18)			   \\
    5  &  3s3p         &    $^3$P$^o_{2}$   &     1.00(  5)				   \\
    6  &  3s3p         &    $^1$P$^o_{1}$   &     0.27(  3)+0.72(  6)			   \\
    7  &  3p$^2$       &    $^1$D$  _{2}$   &     0.28( 17)+0.56(  7)+0.05( 10)+0.12( 12)  \\
    8  &  3p$^2$       &    $^3$P$  _{1}$   &     1.00(  8)				   \\
\hline     
\end{longtable}

                               
\begin{flushleft}                                                                               
                               
{\small
                                                                               
}                                                                                               
                               
\end{flushleft} 

\clearpage
\newpage

\renewcommand{\baselinestretch}{1.0}
\footnotesize
\begin{longtable}{@{\extracolsep\fill}rlllrrr@{}}
\caption{Threshold energies (in Ryd) of the lowest 30 levels of  W LXIV and their lifetimes. ($a{\pm}b \equiv$ $a\times$10$^{{\pm}b}$).}
Index  &     Configuration          & Level        & NIST       & GRASP     &   FAC  &   $\tau$ (s)  \\  \\
 \hline\\
\endfirsthead\\
\caption[]{(continued)}
Index  &     Configuration          & Level        & NIST       & GRASP     &   FAC  &   $\tau$ (s) \\  \\
\hline\\
\endhead 
    1   &   2p$^6$3s	  &     $^2$S$  _{1/2}$    &  000.0   	&   0.0000  &	0.0000   &    ........    \\
    2   &   2p$^6$3p	  &     $^2$P$^o_{1/2}$    &  011.7280	&  11.8989  &  11.7457   &    2.218$-$11  \\
    3   &   2p$^6$3p	  &     $^2$P$^o_{3/2}$    &  039.1890	&  39.3365  &  39.2218   &    5.664$-$13  \\ 
    4   &   2p$^6$3d	  &     $^2$D$  _{3/2}$    &  052.9692	&  53.1127  &  52.9352   &    6.986$-$13  \\
    5   &   2p$^6$3d	  &     $^2$D$  _{5/2}$    &  059.2105	&  59.3372  &  59.1730   &    4.987$-$12  \\
    6   &   2p$^6$4s	  &     $^2$S$  _{1/2}$    &  239.12	& 239.0661  & 238.9973   &    1.501$-$14  \\
    7   &   2p$^6$4p	  &     $^2$P$^o_{1/2}$    &  243.92	& 243.9788  & 243.8505   &    1.267$-$14  \\
    8   &   2p$^6$4p	  &     $^2$P$^o_{3/2}$    &  255.18	& 255.2154  & 255.0981   &    2.010$-$14  \\
\hline            								                	 
\end{longtable}   								   					       
			      							   					       

														       
\begin{flushleft}													       
{\small
NIST:  {\tt http://www.nist.gov/pml/data/asd.cfm} \\						      		      
GRASP: Present results with the {\sc grasp} code from 50  configurations and 1235 levels \\
FAC: Present results with the {\sc fac} code  from  1592 levels \\		       
}															       
\end{flushleft} 

\renewcommand{\baselinestretch}{1.0}
\footnotesize
\begin{longtable}{@{\extracolsep\fill}rllrrrr@{}}
\caption{Comparison of energies (in Ryd) for the common of levels of  W LXV.}  
Index$^a$  &     Configuration        & Level              & NIST       & GRASP        &    FAC    &  MRMP \\  \\
 \hline\\
\endfirsthead\\
\caption[]{(continued)}
Index$^a$  &     Configuration        & Level              & NIST       & GRASP        &    FAC    &  MRMP  \\  \\
\hline\\
\endhead 
    1  &    2s$^2$2p$^6$	  & 	$^1$S$  _{0}$  &     0.000  &	  0.0000   &   0.0000  &    0.0000   \\
    3  &    2s$^2$2p$^5$3s	  & 	$^1$P$^o_{1}$  &   610.640  &	610.2292   & 610.1423  &  610.5354   \\
    9  &    2s$^2$2p$^5$3p	  & 	$^1$S$  _{0}$  &   653.859  &	653.7288   & 653.5037  &  653.7409   \\
   11  &    2s$^2$2p$^5$3d	  & 	$^3$P$^o_{1}$  &   661.507  &	660.9754   & 660.7169  &  661.1325   \\
   17  &    2s$^2$2p$^5$3d	  & 	$^1$P$^o_{1}$  &   670.246  &	670.5722   & 670.2893  &  670.6958   \\
   19  &    2s$^2$2p$^5$3s	  & 	$^3$P$^o_{1}$  &   711.936  &	711.7088   & 711.6628  &  712.0517   \\
   21  &    2s$^2$2p$^5$3p	  & 	$^3$P$  _{0}$  &   726.088  &	725.8494   & 725.6751  &  725.9370   \\
   27  &    2s2p$^6$3p  	  & 	$^3$P$^o_{1}$  &   758.302  &	758.6381   & 758.2086  &  758.3025   \\
   29  &    2s$^2$2p$^5$3d	  & 	$^3$D$^o_{1}$  &   765.027  &	764.8414   & 764.5743  &  764.9308   \\
   33  &    2s2p$^6$3p  	  & 	$^1$P$^o_{1}$  &   786.651  &	787.2457   & 786.8073  &  786.8504   \\
\\  \hline            								                	 
\end{longtable}   								   					       
			      							   					       

														       
\begin{flushleft}													       
{\small
$a$: See Table 7 for definition of all levels \\
NIST:  {\tt http://www.nist.gov/pml/data/asd.cfm} \\						      		      
GRASP: Present results with the {\sc grasp} code from 25  configurations and 157 levels \\
FAC: Present results with the {\sc fac} code  from  1147 levels \\
MRMP: Earlier calculations of Vilkas et al. \cite{mrmp}		       
}															       
\end{flushleft} 

\renewcommand{\baselinestretch}{1.0}
\footnotesize
\begin{longtable}{@{\extracolsep\fill}rllr@{}}
\caption{Eigenvectors (EV) for the lowest 33 levels of W~LXV from the {\sc grasp} code. Numbers outside and inside a bracket correspond to EV \\and the level, respectively. See Table~7 for the definition of all levels.}
Index  & Configuration           & Level              &  Eigenvectors     \\  \\
 \hline\\
\endfirsthead\\
\caption[]{(continued)}
Index  & Configuration           & Level              &  Eigenvectors   \\  \\
\hline\\
\endhead
    1  &  2s$^2$2p$^6$  	&     $^1$S$  _{0}$   &   1.00(  1)					      \\
    2  &  2s$^2$2p$^5$3s	&     $^3$P$^o_{2}$   &   1.00(  2)					      \\
    3  &  2s$^2$2p$^5$3s	&     $^1$P$^o_{1}$   &   0.34( 19)+0.66(  3)				      \\
    4  &  2s$^2$2p$^5$3p	&     $^3$P$  _{1}$   &   0.09( 20)+0.49(  4)+0.31( 25)+0.10(  6)	      \\
    5  &  2s$^2$2p$^5$3p	&     $^3$D$  _{2}$   &   0.50(  5)+0.17(  8)+0.34( 24) 		      \\
    6  &  2s$^2$2p$^5$3p	&     $^1$P$  _{1}$   &   0.08( 20)+0.36( 25)+0.56(  6) 		      \\
    7  &  2s$^2$2p$^5$3p	&     $^3$D$  _{3}$   &   1.00(  7)					      \\
    8  &  2s$^2$2p$^5$3p	&     $^3$P$  _{2}$   &   0.67(  8)+0.34( 24)				      \\
\hline     
\end{longtable}

                               
\begin{flushleft}                                                                               
                               
{\small
                                                                               
}                                                                                               
                               
\end{flushleft} 

\clearpage
\newpage

\renewcommand{\baselinestretch}{1.0}
\footnotesize
\begin{longtable}{@{\extracolsep\fill}rllr@{}}
\caption{Eigenvectors (EV) for the lowest 48 levels of W~LXVI from the {\sc grasp} code. Numbers outside and inside a bracket correspond to EV \\and the level, respectively. See Table~8 for the definition of all levels.}
Index  & Configuration           & Level              &    Eigenvectors   \\  \\
 \hline\\
\endfirsthead\\
\caption[]{(continued)}
Index  & Configuration           & Level              &  Eigenvectors   \\  \\
\hline\\
\endhead
    1  &    2s$^2$2p$^5$	    &   $^2$P$^o_{ 3/2}$     & 1.00(  1)						  	\\
    2  &    2s$^2$2p$^5$	    &   $^2$P$^o_{ 1/2}$     & 1.00(  2)						  	\\
    3  &    2s2p$^6$	            &   $^2$S$  _{ 1/2}$     & 1.00(  3)						  	\\
    4  &    2s$^2$2p$^4$3s	    &   $^4$P$  _{ 5/2}$     & 0.69(  4)+0.31( 28)					  	\\
    5  &    2s$^2$2p$^4$3s	    &   $^2$P$  _{ 3/2}$     & 0.12( 26)+0.56(  5)+0.32( 29)				  	\\
    6  &    2s$^2$2p$^4$3s	    &   $^2$S$  _{ 1/2}$     & 0.23( 86)+0.12( 27)+0.66(  6)				  	\\
    7  &    2s$^2$2p$^4$($^3$P)3p   &   $^4$P$^o_{ 3/2}$     & 0.07( 31)+0.31(  7)+0.18( 92)+0.09( 13)+0.16( 43)+0.17( 33)	\\
    8  &    2s$^2$2p$^4$($^3$P)3p   &   $^2$D$^o_{ 5/2}$     & 0.27( 38)+0.12( 10)+0.29(  8)+0.25( 32)+0.07( 45)	  	\\
\hline     
\end{longtable}

                               
\begin{flushleft}                                                                               
                               
{\small
                                                                               
}                                                                                               
                               
\end{flushleft} 


\clearpage
\newpage

\renewcommand{\baselinestretch}{1.0}
\footnotesize
\begin{longtable}{@{\extracolsep\fill}rrlll@{}}
\caption{Comparison of oscillator strengths ($f$-values, dimensionless) for some transitions of  W LXII. $a{\pm}b \equiv$ $a\times$10$^{{\pm}b}$.  \\See Table~4 for definition of  level indices.}
I  &  J  &   RMBPT    & GRASP     &   R  \\  \\
 \hline\\
 \endfirsthead\\
\caption[]{(continued)}
I  &  J  &   RMBPT    & GRASP     &   R  \\    \\
\hline\\
\endhead 
     1  &  2  &  3.17$-$2  &  3.17$-$2  &  9.8$-$1  \\  
     1  &  4  &  2.07$-$3  &  2.07$-$3  &  1.0$+$0  \\  
     1  &  6  &  1.05$-$1  &  1.05$-$1  &  1.0$+$0  \\  
     1  &  8  &  4.99$-$1  &  4.99$-$1  &  1.0$+$0  \\  
     1  & 20  &  2.07$-$4  &  2.07$-$4  &  1.3$+$0  \\  
     1  & 22  &  1.22$-$4  &  1.22$-$4  &  7.0$-$1  \\  
     2  &  3  &  3.17$-$4  &  3.23$-$4  &  1.7$+$0  \\  
     3  &  4  &  2.48$-$3  &  2.48$-$3  &  1.0$+$0  \\
\hline            								                	 
\end{longtable}   								   					       
			      							   					       

														       
\begin{flushleft}													       
{\small
RMBPT: Earlier results of Safronova and Safronova \cite{saf} \\
GRASP: Present results with the {\sc grasp} code from 63  configurations and 2003 levels \\
R: Ratio of velocity and lrength forms of $f$-values \\
}															       
\end{flushleft} 

\clearpage
\newpage

\renewcommand{\baselinestretch}{1.0}
\footnotesize
\begin{longtable}{@{\extracolsep\fill}rrlll@{}}
\caption{Comparison of oscillator strengths ($f$-values, dimensionless) for some transitions of  W LXIII. $a{\pm}b \equiv$ $a\times$10$^{{\pm}b}$. \\See Table~5 for definition of  level indices.}
    I  &  J  &   RMBPT    & GRASP     &   R \\  \\
 \hline\\
\endfirsthead\\
\caption[]{(continued)}
    I  &  J  &   RMBPT    & GRASP     &   R         \\  \\
\hline\\
\endhead 
     1  &  6 & 5.97$-$1 &   6.08$-$1  &   1.0$+$0  \\
     2  &  8 & 2.76$-$1 &   2.81$-$1  &   1.0$+$0  \\
     2  &  9 & 2.23$-$1 &   2.27$-$1  &   1.0$+$0  \\
     3  &  4 & 3.11$-$2 &   3.18$-$2  &   1.0$+$0  \\
     3  &  7 & 6.43$-$2 &   6.45$-$2  &   1.0$+$0  \\
     3  &  8 & 5.27$-$2 &   5.36$-$2  &   1.0$+$0  \\
     3  &  9 & 3.61$-$2 &   3.67$-$2  &   1.0$+$0  \\
     3  & 10 & 3.12$-$1 &   3.19$-$1  &   1.0$+$0  \\
\hline            								                	 
\end{longtable}   								   					       
			      							   					       

														       
\begin{flushleft}													       
{\small
RMBPT: Earlier results of Safronova and Safronova \cite{saf} \\
GRASP: Present results with the {\sc grasp} code from 58  configurations and 509 levels \\
R: Ratio of velocity and length forms of $f$-values \\		       
}															       
\end{flushleft} 


\clearpage
\newpage

\renewcommand{\baselinestretch}{1.0}
\footnotesize
\begin{longtable}{@{\extracolsep\fill}rrrrrr@{}}
\caption{ Comparison of radiative rates ($A$-values, s$^{-1}$) for some transitions of  W LXV. $a{\pm}b \equiv$ $a\times$10$^{{\pm}b}$. \\See Table~7 for definition of  level indices.} 
   I  &  J  &   MRMP    & GRASP     &  f (GRASP) &  R     \\  \\
 \hline\\
\endfirsthead\\
\caption[]{(continued)}
     I  &  J  &   MRMP    & GRASP     &   f (GRASP) &  R                  \\  \\
\hline\\
\endhead 
     1  &  3 & 1.206$+$14 & 1.5309$+$14 & 1.5354$-$1  &  1.0$-$0  \\  
     1  & 11 & 6.551$+$13 & 8.2270$+$13 & 7.0330$-$2  &  9.8$-$1  \\  
     1  & 17 & 2.613$+$15 & 2.8077$+$15 & 2.3320$+$0  &  9.8$-$1  \\  
     1  & 19 & 2.694$+$13 & 3.8180$+$13 & 2.8152$-$2  &  9.9$-$1  \\  
     1  & 27 & 6.243$+$14 & 7.7623$+$14 & 5.0372$-$1  &  1.0$-$0  \\  
     1  & 29 & 1.227$+$15 & 1.3590$+$15 & 8.6763$-$1  &  9.8$-$1  \\  
     1  & 33 & 3.350$+$14 & 4.4021$+$14 & 2.6529$-$1  &  1.0$-$0  \\  
     1  & 39 & 4.193$+$13 & 5.3006$+$13 & 2.7134$-$2  &  9.6$-$1  \\
\hline            								                	 
\end{longtable}   								   					       
			      							   					       

														       
\begin{flushleft}													       
{\small
MRMP: Earlier results of Vilkas et al. \cite{mrmp}  \\
GRASP: Present results with the {\sc grasp} code from 25  configurations and 157 levels \\
R: Ratio of velocity and lrength forms of $f$-values \\		       
}															       
\end{flushleft} 


\clearpage
\newpage


\TableExplanation

\bigskip
\renewcommand{\arraystretch}{1.0}

\section*{Table 1.\label{tbl1te} Energies (Ryd) for the lowest 220 levels of W~LIX and their lifetimes ($\tau$, s). }
\begin{tabular}{@{}p{1in}p{6in}@{}}
Index            & Level Index \\
Configuration    & The configuration to which the level belongs \\
Level             & The $LSJ$ designation of the level \\
GRASP          & Present energies from the {\sc grasp} code  with 12~652 level calculations \\
FAC             & Present energies from the {\sc fac} code  with 38~694 level calculations \\
$\tau$ (s)       & Lifetime of the level in s \\

\end{tabular}
\label{tableII}

\bigskip
\renewcommand{\arraystretch}{1.0}

\section*{Table 2.\label{tbl2te} Energies (Ryd) for the lowest 220 levels of W~LX and their lifetimes ($\tau$, s). }
\begin{tabular}{@{}p{1in}p{6in}@{}}
Index            & Level Index \\
Configuration    & The configuration to which the level belongs \\
Level             & The $LSJ$ designation of the level \\
GRASP          & Present energies from the {\sc grasp} code  with 3533 level calculations \\
FAC             & Present energies from the {\sc fac} code  with 14~608 level calculations \\
$\tau$ (s)       & Lifetime of the level in s \\

\end{tabular}
\label{tableII}

\bigskip
\renewcommand{\arraystretch}{1.0}

\section*{Table 3.\label{tbl3te} Energies (Ryd) for the lowest 215 levels of W~LXI and their lifetimes ($\tau$, s). }
\begin{tabular}{@{}p{1in}p{6in}@{}}
Index            & Level Index \\
Configuration    & The configuration to which the level belongs \\
Level             & The $LSJ$ designation of the level \\
GRASP          & Present energies from the {\sc grasp} code  with 4364 level calculations \\
FAC             & Present energies from the {\sc fac} code  with 27~122 level calculations \\
$\tau$ (s)       & Lifetime of the level in s \\

\end{tabular}
\label{tableII}

\bigskip
\renewcommand{\arraystretch}{1.0}

\section*{Table 4.\label{tbl4te} Energies (Ryd) for the lowest 148 levels of W~LXII and their lifetimes ($\tau$, s). }
\begin{tabular}{@{}p{1in}p{6in}@{}}
Index            & Level Index \\
Configuration    & The configuration to which the level belongs \\
Level             & The $LSJ$ designation of the level \\
GRASP          & Present energies from the {\sc grasp} code  with 2003 level calculations \\
FAC             & Present energies from the {\sc fac} code  with 12~139 level calculations \\
$\tau$ (s)       & Lifetime of the level in s \\

\end{tabular}
\label{tableII}

\bigskip
\renewcommand{\arraystretch}{1.0}

\section*{Table 5.\label{tbl5te} Energies (Ryd) for the lowest 210 levels of W~LXIII and their lifetimes ($\tau$, s). }
\begin{tabular}{@{}p{1in}p{6in}@{}}
Index            & Level Index \\
Configuration    & The configuration to which the level belongs \\
Level             & The $LSJ$ designation of the level \\
GRASP          & Present energies from the {\sc grasp} code  with 509 level calculations \\
FAC             & Present energies from the {\sc fac} code  with 991 level calculations \\
$\tau$ (s)       & Lifetime of the level in s \\

\end{tabular}
\label{tableII}

\bigskip
\renewcommand{\arraystretch}{1.0}

\section*{Table 6.\label{tbl6te} Energies (Ryd) for the 2p$^5$$n\ell$ ($n \le$ 20)  levels of W~LXIV. }
\begin{tabular}{@{}p{1in}p{6in}@{}}
Index            & Level Index \\
Configuration    & The configuration to which the level belongs \\
Level             & The $LSJ$ designation of the level \\
FAC             & Present energies from the {\sc fac} code  with 1592 level calculations \\
\end{tabular}
\label{tableII}

\bigskip
\renewcommand{\arraystretch}{1.0}

\section*{Table 7.\label{tbl7te} Energies (Ryd) for the lowest 121 levels of W~LXV and their lifetimes ($\tau$, s). }
\begin{tabular}{@{}p{1in}p{6in}@{}}
Index            & Level Index \\
Configuration    & The configuration to which the level belongs \\
Level             & The $LSJ$ designation of the level \\
GRASP          & Present energies from the {\sc grasp} code  with 157 level calculations \\
FAC             & Present energies from the {\sc fac} code  with 1147 level calculations \\
MRMP         & Earlier energies of Vilas et al. \cite{mrmp} \\ 
$\tau$ (GRASP, s)       & Lifetime  of the level in s from the GRASP calculations  \\
$\tau$ (MRMP, s)       & Lifetime  of the level in s from the Vilkas et al. \cite{mrmp}   \\
\end{tabular}
\label{tableII}

\bigskip
\renewcommand{\arraystretch}{1.0}

\section*{Table 8.\label{tbl8te} Energies (Ryd) for the lowest 150 levels of W~LXVI and their lifetimes ($\tau$, s). }
\begin{tabular}{@{}p{1in}p{6in}@{}}
Index            & Level Index \\
Configuration    & The configuration to which the level belongs \\
Level             & The $LSJ$ designation of the level \\
GRASP          & Present energies from the {\sc grasp} code  with 501 level calculations \\
FAC             & Present energies from the {\sc fac} code  with 1113 level calculations \\
$\tau$ (s)       & Lifetime of the level in s \\

\end{tabular}
\label{tableII}


\bigskip
\section*{Table 9.\label{tbl9te}  Transition wavelengths ($\lambda_{ij}$ in $\rm \AA$), radiative rates ($A_{ji}$ in s$^{-1}$),
 oscillator strengths ($f_{ij}$, dimensionless), and line strengths ($S$, in atomic units) for electric dipole (E1), and 
$A_{ji}$ for electric quadrupole (E2), magnetic dipole (M1), and magnetic quadrupole (M2) transitions of W~LIX.
 The ratio R(E1) of velocity and length forms of $A$-values for E1 transitions is listed in the last column.}
\begin{tabular}{@{}p{1in}p{6in}@{}}
$i$ and $j$         & The lower ($i$) and upper ($j$) levels of a transition as defined in Table 1.\\
$\lambda_{ij}$      & Transition wavelength (in ${\rm \AA}$) \\
$A^{E1}_{ji}$       & Radiative transition probability (in s$^{-1}$) for the E1 transitions \\
$f^{E1}_{ij}$       & Absorption oscillator strength (dimensionless) for the E1 transitions \\
$S^{E1}$            & Line strength in atomic unit (a.u.), 1 a.u. = 6.460$\times$10$^{-36}$ cm$^2$ esu$^2$ for the E1 transitions \\
$A^{E2}_{ji}$       & Radiative transition probability (in s$^{-1}$) for the E2 transitions \\
$A^{M1}_{ji}$       & Radiative transition probability (in s$^{-1}$) for the M1 transitions \\
$A^{M2}_{ji}$       & Radiative transition probability (in s$^{-1}$) for the M2 transitions \\
R(E1)                     & Ratio of velocity and length forms of $A$- (or $f$- and $S$-) values for the E1 transitions \\
$a{\pm}b$ &  $\equiv a\times{10^{{\pm}b}}$ \\
\end{tabular}
\label{ExplTable9}

\bigskip
\section*{Table 10.\label{tbl10te}  Transition wavelengths ($\lambda_{ij}$ in $\rm \AA$), radiative rates ($A_{ji}$ in s$^{-1}$),
 oscillator strengths ($f_{ij}$, dimensionless), and line strengths ($S$, in atomic units) for electric dipole (E1), and 
$A_{ji}$ for electric quadrupole (E2), magnetic dipole (M1), and magnetic quadrupole (M2) transitions of W~LX.
The ratio R(E1) of velocity and length forms of $A$-values for E1 transitions is listed in the last column.}
\begin{tabular}{@{}p{1in}p{6in}@{}}
$i$ and $j$         & The lower ($i$) and upper ($j$) levels of a transition as defined in Table 2.\\
$\lambda_{ij}$      & Transition wavelength (in ${\rm \AA}$) \\
$A^{E1}_{ji}$       & Radiative transition probability (in s$^{-1}$) for the E1 transitions \\
$f^{E1}_{ij}$       & Absorption oscillator strength (dimensionless) for the E1 transitions \\
$S^{E1}$            & Line strength in atomic unit (a.u.), 1 a.u. = 6.460$\times$10$^{-36}$ cm$^2$ esu$^2$ for the E1 transitions \\
$A^{E2}_{ji}$       & Radiative transition probability (in s$^{-1}$) for the E2 transitions \\
$A^{M1}_{ji}$       & Radiative transition probability (in s$^{-1}$) for the M1 transitions \\
$A^{M2}_{ji}$       & Radiative transition probability (in s$^{-1}$) for the M2 transitions \\
R(E1)                     & Ratio of velocity and length forms of $A$- (or $f$- and $S$-) values for the E1 transitions \\
$a{\pm}b$ &  $\equiv a\times{10^{{\pm}b}}$ \\
\end{tabular}
\label{ExplTable10}

\bigskip
\section*{Table 11.\label{tbl11te}  Transition wavelengths ($\lambda_{ij}$ in $\rm \AA$), radiative rates ($A_{ji}$ in s$^{-1}$),
 oscillator strengths ($f_{ij}$, dimensionless), and line strengths ($S$, in atomic units) for electric dipole (E1), and 
$A_{ji}$ for electric quadrupole (E2), magnetic dipole (M1), and magnetic quadrupole (M2) transitions of W~LXI.
 The ratio R(E1) of velocity and length forms of $A$-values for E1 transitions is listed in the last column.}
\begin{tabular}{@{}p{1in}p{6in}@{}}
$i$ and $j$         & The lower ($i$) and upper ($j$) levels of a transition as defined in Table 3.\\
$\lambda_{ij}$      & Transition wavelength (in ${\rm \AA}$) \\
$A^{E1}_{ji}$       & Radiative transition probability (in s$^{-1}$) for the E1 transitions \\
$f^{E1}_{ij}$       & Absorption oscillator strength (dimensionless) for the E1 transitions \\
$S^{E1}$            & Line strength in atomic unit (a.u.), 1 a.u. = 6.460$\times$10$^{-36}$ cm$^2$ esu$^2$ for the E1 transitions \\
$A^{E2}_{ji}$       & Radiative transition probability (in s$^{-1}$) for the E2 transitions \\
$A^{M1}_{ji}$       & Radiative transition probability (in s$^{-1}$) for the M1 transitions \\
$A^{M2}_{ji}$       & Radiative transition probability (in s$^{-1}$) for the M2 transitions \\
R(E1)                     & Ratio of velocity and length forms of $A$- (or $f$- and $S$-) values for the E1 transitions \\
$a{\pm}b$ &  $\equiv a\times{10^{{\pm}b}}$ \\
\end{tabular}
\label{ExplTable11}

\bigskip
\section*{Table 12.\label{tbl12te}  Transition wavelengths ($\lambda_{ij}$ in $\rm \AA$), radiative rates ($A_{ji}$ in s$^{-1}$),
 oscillator strengths ($f_{ij}$, dimensionless), and line strengths ($S$, in atomic units) for electric dipole (E1), and 
$A_{ji}$ for electric quadrupole (E2), magnetic dipole (M1), and magnetic quadrupole (M2) transitions of W~LXII.
 The ratio R(E1) of velocity and length forms of $A$-values for E1 transitions is listed in the last column.}
\begin{tabular}{@{}p{1in}p{6in}@{}}
$i$ and $j$         & The lower ($i$) and upper ($j$) levels of a transition as defined in Table 4.\\
$\lambda_{ij}$      & Transition wavelength (in ${\rm \AA}$) \\
$A^{E1}_{ji}$       & Radiative transition probability (in s$^{-1}$) for the E1 transitions \\
$f^{E1}_{ij}$       & Absorption oscillator strength (dimensionless) for the E1 transitions \\
$S^{E1}$            & Line strength in atomic unit (a.u.), 1 a.u. = 6.460$\times$10$^{-36}$ cm$^2$ esu$^2$ for the E1 transitions \\
$A^{E2}_{ji}$       & Radiative transition probability (in s$^{-1}$) for the E2 transitions \\
$A^{M1}_{ji}$       & Radiative transition probability (in s$^{-1}$) for the M1 transitions \\
$A^{M2}_{ji}$       & Radiative transition probability (in s$^{-1}$) for the M2 transitions \\
R(E1)                     & Ratio of velocity and length forms of $A$- (or $f$- and $S$-) values for the E1 transitions \\
$a{\pm}b$ &  $\equiv a\times{10^{{\pm}b}}$ \\
\end{tabular}
\label{ExplTable12}

\bigskip
\section*{Table 13.\label{tbl13te}  Transition wavelengths ($\lambda_{ij}$ in $\rm \AA$), radiative rates ($A_{ji}$ in s$^{-1}$),
 oscillator strengths ($f_{ij}$, dimensionless), and line strengths ($S$, in atomic units) for electric dipole (E1), and 
$A_{ji}$ for electric quadrupole (E2), magnetic dipole (M1), and magnetic quadrupole (M2) transitions of W~LXIII.
 The ratio R(E1) of velocity and length forms of $A$-values for E1 transitions is listed in the last column.}
\begin{tabular}{@{}p{1in}p{6in}@{}}
$i$ and $j$         & The lower ($i$) and upper ($j$) levels of a transition as defined in Table 5.\\
$\lambda_{ij}$      & Transition wavelength (in ${\rm \AA}$) \\
$A^{E1}_{ji}$       & Radiative transition probability (in s$^{-1}$) for the E1 transitions \\
$f^{E1}_{ij}$       & Absorption oscillator strength (dimensionless) for the E1 transitions \\
$S^{E1}$            & Line strength in atomic unit (a.u.), 1 a.u. = 6.460$\times$10$^{-36}$ cm$^2$ esu$^2$ for the E1 transitions \\
$A^{E2}_{ji}$       & Radiative transition probability (in s$^{-1}$) for the E2 transitions \\
$A^{M1}_{ji}$       & Radiative transition probability (in s$^{-1}$) for the M1 transitions \\
$A^{M2}_{ji}$       & Radiative transition probability (in s$^{-1}$) for the M2 transitions \\
R(E1)                     & Ratio of velocity and length forms of $A$- (or $f$- and $S$-) values for the E1 transitions \\
$a{\pm}b$ &  $\equiv a\times{10^{{\pm}b}}$ \\
\end{tabular}
\label{ExplTable13}

\bigskip
\section*{Table 14.\label{tbl14te}  Transition wavelengths ($\lambda_{ij}$ in $\rm \AA$), radiative rates ($A_{ji}$ in s$^{-1}$),
 oscillator strengths ($f_{ij}$, dimensionless), and line strengths ($S$, in atomic units) for electric dipole (E1), and 
$A_{ji}$ for electric quadrupole (E2), magnetic dipole (M1), and magnetic quadrupole (M2) transitions of W~LXIV.
  The ratio R(E1) of velocity and length forms of $A$-values for E1 transitions is listed in the last column.}
\begin{tabular}{@{}p{1in}p{6in}@{}}
$i$ and $j$         & The lower ($i$) and upper ($j$) levels of a transition as defined in Table 6.\\
$\lambda_{ij}$      & Transition wavelength (in ${\rm \AA}$) \\
$A^{E1}_{ji}$       & Radiative transition probability (in s$^{-1}$) for the E1 transitions \\
$f^{E1}_{ij}$       & Absorption oscillator strength (dimensionless) for the E1 transitions \\
$S^{E1}$            & Line strength in atomic unit (a.u.), 1 a.u. = 6.460$\times$10$^{-36}$ cm$^2$ esu$^2$ for the E1 transitions \\
$A^{E2}_{ji}$       & Radiative transition probability (in s$^{-1}$) for the E2 transitions \\
$A^{M1}_{ji}$       & Radiative transition probability (in s$^{-1}$) for the M1 transitions \\
$A^{M2}_{ji}$       & Radiative transition probability (in s$^{-1}$) for the M2 transitions \\
R(E1)                     & Ratio of velocity and length forms of $A$- (or $f$- and $S$-) values for the E1 transitions \\
$a{\pm}b$ &  $\equiv a\times{10^{{\pm}b}}$ \\
\end{tabular}
\label{ExplTable14}

\bigskip
\section*{Table 15.\label{tbl15te}  Transition wavelengths ($\lambda_{ij}$ in $\rm \AA$), radiative rates ($A_{ji}$ in s$^{-1}$),
 oscillator strengths ($f_{ij}$, dimensionless), and line strengths ($S$, in atomic units) for electric dipole (E1), and 
$A_{ji}$ for electric quadrupole (E2), magnetic dipole (M1), and magnetic quadrupole (M2) transitions of W~LXV.
  The ratio R(E1) of velocity and length forms of $A$-values for E1 transitions is listed in the last column.}
\begin{tabular}{@{}p{1in}p{6in}@{}}
$i$ and $j$         & The lower ($i$) and upper ($j$) levels of a transition as defined in Table 7.\\
$\lambda_{ij}$      & Transition wavelength (in ${\rm \AA}$) \\
$A^{E1}_{ji}$       & Radiative transition probability (in s$^{-1}$) for the E1 transitions \\
$f^{E1}_{ij}$       & Absorption oscillator strength (dimensionless) for the E1 transitions \\
$S^{E1}$            & Line strength in atomic unit (a.u.), 1 a.u. = 6.460$\times$10$^{-36}$ cm$^2$ esu$^2$ for the E1 transitions \\
$A^{E2}_{ji}$       & Radiative transition probability (in s$^{-1}$) for the E2 transitions \\
$A^{M1}_{ji}$       & Radiative transition probability (in s$^{-1}$) for the M1 transitions \\
$A^{M2}_{ji}$       & Radiative transition probability (in s$^{-1}$) for the M2 transitions \\
R(E1)                     & Ratio of velocity and length forms of $A$- (or $f$- and $S$-) values for the E1 transitions \\
$a{\pm}b$ &  $\equiv a\times{10^{{\pm}b}}$ \\
\end{tabular}
\label{ExplTable15}

\bigskip
\section*{Table 16.\label{tbl16te}  Transition wavelengths ($\lambda_{ij}$ in $\rm \AA$), radiative rates ($A_{ji}$ in s$^{-1}$),
 oscillator strengths ($f_{ij}$, dimensionless), and line strengths ($S$, in atomic units) for electric dipole (E1), and 
$A_{ji}$ for electric quadrupole (E2), magnetic dipole (M1), and magnetic quadrupole (M2) transitions of W~LXVI.
  The ratio R(E1) of velocity and length forms of $A$-values for E1 transitions is listed in the last column.}
\begin{tabular}{@{}p{1in}p{6in}@{}}
$i$ and $j$         & The lower ($i$) and upper ($j$) levels of a transition as defined in Table 8.\\
$\lambda_{ij}$      & Transition wavelength (in ${\rm \AA}$) \\
$A^{E1}_{ji}$       & Radiative transition probability (in s$^{-1}$) for the E1 transitions \\
$f^{E1}_{ij}$       & Absorption oscillator strength (dimensionless) for the E1 transitions \\
$S^{E1}$            & Line strength in atomic unit (a.u.), 1 a.u. = 6.460$\times$10$^{-36}$ cm$^2$ esu$^2$ for the E1 transitions \\
$A^{E2}_{ji}$       & Radiative transition probability (in s$^{-1}$) for the E2 transitions \\
$A^{M1}_{ji}$       & Radiative transition probability (in s$^{-1}$) for the M1 transitions \\
$A^{M2}_{ji}$       & Radiative transition probability (in s$^{-1}$) for the M2 transitions \\
R(E1)                     & Ratio of velocity and length forms of $A$- (or $f$- and $S$-) values for the E1 transitions \\
$a{\pm}b$ &  $\equiv a\times{10^{{\pm}b}}$ \\
\end{tabular}
\label{ExplTable16}

\datatables 


\end{document}